\documentclass[aps,prd,onecolumn,groupedaddress,showpacs,nofootinbib,amssymb]{revtex4}
\usepackage[dvips]{graphicx}
\usepackage{amssymb}
\usepackage{amsmath}
\usepackage{graphicx,,color}
\usepackage{amsfonts}
\usepackage{bm}
\usepackage{cancel}
\usepackage{comment}

\newcommand\be{\begin{equation}}
\newcommand\ee{\end{equation}}

\allowdisplaybreaks[4]

\begin{document}

\title{Scalar Field Assisted $f(R)$ Gravity Inflation}
\author{
K. Kleidis,$^{1}$\,\thanks{kleidis@teiser.gr}
V.K. Oikonomou$^{2,1}$,\,\thanks{v.k.oikonomou1979@gmail.com}}
\affiliation{
$^{1)}$ Department of Mechanical Engineering\\ Technological
Education Institute of Central Macedonia \\
62124 Serres, Greece
}
\affiliation{
$^{2)}$Department of Physics, Aristotle University of Thessaloniki, Thessaloniki 54124, Greece}

\tolerance=5000

\begin{abstract}
In this paper we investigate the inflationary dynamics of an $f(R)$
gravity in the presence of a canonical scalar field. We specifically
choose the cosmological evolution to be a quasi-de Sitter evolution
and also the $f(R)$ gravity is assumed to be a modified version of
the $R^2$ gravity. We investigate which scalar field potential can
produce the quasi-de Sitter evolution for the choice of the $f(R)$
gravity we made, and also we study in detail the inflationary
dynamics of the resulting theory. As we demonstrate, the spectral
index is identical to the one corresponding to the ordinary $R^2$
gravity, while the scalar-to-tensor ratio is found to be smaller
than the $R^2$ inflation one consequently, compatibility with both
the Planck 2015 and BICEP2/Keck-Array data is achieved.
\end{abstract}


\maketitle

\section{Introduction}

One of the most appealing scenarios of the early Universe is the
inflationary scenario, which describes the abrupt nearly exponential
accelerating expansion of the Universe in a quite small time
interval. In modern theoretical cosmology, two of the most prominent
trends in describing this early-time acceleration are the scalar
field description \cite{Linde:2007fr,Gorbunov:2011zzc,Lyth:1998xn}
and the modified gravity description of inflation
\cite{Nojiri:2006ri,Capozziello:2011et,Capozziello:2010zz,
Capozziello:2009nq,Nojiri:2010wj,Nojiri:2017ncd,Clifton:2011jh}. In
both cases, various models of inflation were proposed from time to
time, however after the latest Planck collaboration observational
constraints \cite{Ade:2015lrj}, many models of inflation were
disproved or severely constrained, see \cite{Martin:2016ckm} for a
recent review on this issue. In general there are various
scalar-tensor models which remain viable even after the Planck data
were released, for example the Starobinsky model
\cite{Starobinsky:1980te,Barrow:1988xh}, the Higgs model
\cite{Bezrukov:2007ep} and a recently discovered class of scalar
exponential models called $\alpha$-attractors
\cite{Kallosh:2013hoa,Ferrara:2013rsa,Kallosh:2013yoa}. Also, among
various modified gravity theories, the $f(R)$ gravity models combine
an elegant simplicity with the successful confrontation with the
Planck data, and also for the $f(R)$ gravity description of the
$\alpha$-attractor models, see
\cite{Odintsov:2016vzz,Odintsov:2016jwr}.

In view of the appealing properties of $f(R)$ gravity and of the
canonical scalar field inflationary models, in this paper we shall
consider $f(R)$ gravity in the presence of a canonical scalar field.
This theory is the simplest form of an $f(R,\phi)$ inflationary
theory \cite{Nojiri:2017ncd}, so in this work we shall investigate
how a scalar field affects the viability of a specifically chosen
$f(R)$ gravity. We shall assume that the $f(R)$ gravity is a
modified version of the $R^2$ model and we shall study the effects
and implications of the scalar field on the observational indices.
Our method is simple, since we will fix the Hubble rate to be a
quasi-de Sitter evolution, and we shall find the form of the
potential for which such an evolution is possible. Our main
assumption for the $f(R)$ gravity sector is that the slow-roll
approximation holds true, and the same applies for the scalar field
sector. With the present study we aim to provide a scalar-tensor
assisted modified $R^2$ gravity model, which as we shall demonstrate
remains viable, for a wider range of the parameter values, when the
scalar-to-tensor ratio is considered.

The reasons for choosing a modified version of the $R^2$ gravity are
firstly that any version of the $R^2$ gravity is functionally
simple, and secondly the ordinary $R^2$ gravity yields
inconsistencies when the term $\dot{\phi}^2$ is considered, as we
also discuss in the main text.

Thus in the next section, we study the modified $R^2$ inflationary
scenario in the presence of a slow-rolling canonical scalar field.
We realize a quasi-de Sitter Hubble rate, and we investigate which
potential can assist the $f(R)$ gravity to realize such an
evolution. Also we calculate in detail the slow-roll indices and the
corresponding observational indices and we confront the model with
the observational data.

Before we proceed to the core of our article, we briefly present the
geometric background we shall assume. Firstly, the metric will be
assumed to be the flat  Friedmann-Robertson-Walker (FRW), the line
element of which is,
\begin{equation}
\label{metricfrw} ds^2 = - dt^2 + a(t)^2 \sum_{i=1,2,3}
\left(dx^i\right)^2\, ,
\end{equation}
with $a(t)$ being the scale factor. In addition, the metric
connection is the Levi-Civita connection, which is torsion-less,
metric compatible and symmetric.

\section{Slow-roll Scalar Field Assisted Modified $R^2$ Gravity}

The theoretical framework which we shall consider in this article,
is described by the simplest $f(R,\phi)$ gravity, since it is an
$f(R)$ gravity in the presence of a canonical scalar field, with the
action being,
\begin{equation}
\label{JGRG7} \mathcal{S}=\int d ^4x\sqrt{-g}\left
(\frac{f(R)}{2\kappa^2} -\frac{1}{2}
g^{\mu\nu}\partial_{\mu}\phi\partial_{\nu}\phi-V(\phi )\right)\, ,
\end{equation}
and for simplicity we shall use a physical units system for which
$\kappa^2=1$. By varying the gravitational action (\ref{JGRG7}),
with respect to the metric tensor, we obtain the cosmological
equations of motion,
\begin{align}
\label{eqnmotion1} & 3FH^2= \frac{FR-f}{2}-3H\dot{F}+
\frac{\dot{\phi}^2}{2}+V(\phi)\, , \\ & \label{eqnmotion2}
-2F\dot{H}= \ddot{F}-H\dot{F}+\dot{\phi}^2 \, ,
\end{align}
with $F=\frac{\partial f}{\partial R}$ and the ``dot'' indicates as
usual, differentiation with respect to the cosmic time $t$. Also
upon variation of the action (\ref{JGRG7}) with respect to the
scalar field $\phi$, we obtain,
\begin{equation}\label{eqnmotion1121}
\ddot{\phi}+3H\dot{\phi}+\frac{\mathrm{d}V}{\mathrm{d}\phi}=0\, .
\end{equation}
Our main interest in this paper is the inflationary dynamics, and
the quantification of the dynamics is achieved by using the
slow-roll indices $\epsilon_i$, $i=1,...4$, which are defined as
follows
\cite{Noh:2001ia,Hwang:2001qk,Hwang:2001pu,Nojiri:2016vhu,Odintsov:2016plw,Odintsov:2015gba,Nojiri:2017ncd},
\begin{equation}
\label{slowrollgenerarlfrphi} \epsilon_1=-\frac{\dot{H}}{H^2}\,
,\quad \epsilon_2=\frac{\ddot{\phi}}{H\dot{\phi}}\, , \quad
\epsilon_3=\frac{\dot{F}}{2HF}\, ,\quad
\epsilon_4=\frac{\dot{E}}{2HE}\, ,
\end{equation}
where the function $E$ appearing in
Eq.~(\ref{slowrollgenerarlfrphi}) is defined as follows,
\begin{equation}
\label{epsilonfnction} E=F+\frac{3\dot{F}^2}{2\dot{\phi}^2}\, .
\end{equation}
Also at this point we introduce the function $Q_s$, the analytic
form of which is,
\begin{equation}
\label{qsfunction}
Q_s=\dot{\phi}^2\frac{E}{F\,H^2(1+\epsilon_3)^2}\, ,
\end{equation}
which will play a prominent role in the calculation of the
scalar-to-tensor ratio.

Let us now assume that the functional form of the $f(R)$ gravity is
that of a modified $R^2$ model, that is,
\begin{equation}\label{rsquaremodelfr}
f(R)=R-\alpha R^2\, ,
\end{equation}
where $\alpha$ is for the moment a free parameter and also we shall
assume that the Hubble rate $H$ is described by a quasi-de Sitter
evolution,
\begin{equation}\label{hubblerate}
H(t)=H_0-H_it\, .
\end{equation}
The reason for choosing the modified version of the $R^2$ gravity,
is the functional simplicity, but also there is another reason, as
we discuss later on. In addition, both the $f(R)$ gravity and the
Hubble rate can be freely chosen, but this does not ensure that a
viable inflationary era can be obtained. As we show later on, the
choices (\ref{rsquaremodelfr}) and (\ref{hubblerate}) result to a
viable inflationary era. Then, by using the gravitational equations
(\ref{eqnmotion1}), (\ref{eqnmotion2}) and (\ref{eqnmotion1121}), we
can find an approximate form for the scalar potential at leading
order. Indeed, the potential as a function of the cosmic time is
equal to,
\begin{align}
\label{potentialgeneral} & V(\phi (t))=- \frac{FR-f}{2}+3H\dot{F}
-\frac{H\dot{F}-2F\dot{H}- \ddot{F}}{2}+3FH^2\, .
\end{align}
So by substituting the functional form of $f(R)$ and $H(t)$ from
Eqs. (\ref{rsquaremodelfr}) and (\ref{hubblerate}) and also taking
into account that the scalar curvature for a FRW metric, during the
slow-roll era is $R\simeq 12 H^2$, the scalar potential at leading
order in the cosmic time is,
\begin{equation}\label{timepotential}
V(t)=6 H_0^2-12 H_0 H_i t+3 H_0-24 \alpha H_i^2+6 H_i^2 t^2-3 H_i
t\, .
\end{equation}
From Eq. (\ref{eqnmotion2}), by integrating the square root of it,
we can find the function $\phi (t)$, which is,
\begin{equation}\label{phioft}
\phi (t)=4 \sqrt{3 \alpha } H_i t \, .
\end{equation}
At this point we shall explain the choice (\ref{rsquaremodelfr}) in
some detail. Actually, if we chose the ordinary $f(R)$ gravity
model, this would result to $\dot{\phi}^2=-48 \alpha  H_i^2$, which
means that $\phi (t)$ is not physical. Hence, the choice
(\ref{rsquaremodelfr}) was the simplest choice which would could
make, and as we will show, it yields a viable inflationary
cosmology, always in the presence of a canonical scalar field.

Let us proceed to the calculation of the potential, so by inverting
(\ref{phioft}) and substituting in Eq. (\ref{timepotential}), we
obtain $V(\phi)$, which is,
\begin{equation}\label{potentialphispecificform}
V(\phi)=\frac{\phi ^2}{8 \alpha }-\frac{\sqrt{3} \phi }{4 \alpha }+6
H_0^2-\frac{\sqrt{3} H_0 \phi }{\sqrt{\alpha }}+3 H_0-24 \alpha
H_i^2\, .
\end{equation}
So far the parameter $\alpha$ appearing in the functional form of
the $R^2$ gravity, namely Eq. (\ref{rsquaremodelfr}), is left
unidentified, however this will be fixed by requiring Eq.
(\ref{eqnmotion1121}) to hold true for the solution (\ref{phioft})
and for the potential (\ref{potentialphispecificform}). Since we
assume that the scalar field is also slowly rolling down its
potential, the term $\ddot{\phi}$ can be omitted in Eq.
(\ref{eqnmotion1121}), but in any case, from Eq. (\ref{phioft}), we
can easily see that it is equal to zero. Hence the condition
$3H\dot{\phi}+V'(\phi)$, fixes the value of $\alpha$ at leading
order, which is,
\begin{equation}\label{alphavalues1}
\alpha=\frac{16 H_0^2+8 H_0+1}{48 H_0^2}\, .
\end{equation}

\begin{figure}[h]
\centering
\includegraphics[width=18pc]{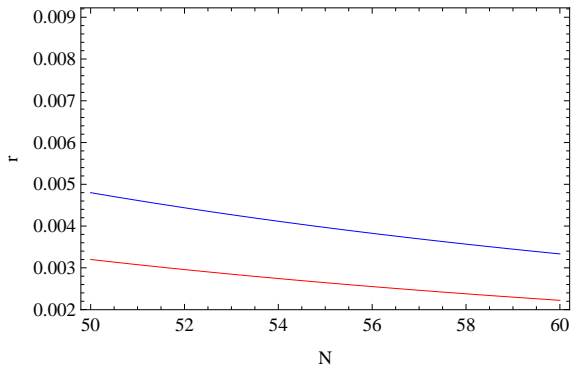}
\caption{The scalar-to-tensor ratio $r=\frac{8}{N^2}$ (red curve)
and $r=\frac{12}{N^2}$ (blue curve), as a function of the
$e$-foldings number $N$.}\label{plot2}
\end{figure}

Now we turn our focus to the inflationary dynamics of the $f(R)$
gravity in the presence of the scalar field. As we already
indicated, we assume that the slow-roll approximation holds true in
the $f(R)$ gravity sector, and also the scalar field slowly rolls
down its potential. In effect we expect that the second slow-roll
index, namely $\epsilon_2$ in Eq. (\ref{slowrollgenerarlfrphi}), is
approximately zero. Remarkably though, a direct calculation results
to the following equation,
\begin{equation}\label{epsilon2general}
\epsilon_2=-\frac{H\ddot{F}-\frac{\mathrm{d}^2F}{\mathrm{d}t^3}-2F\ddot{H}-\dot{H}\dot{F}}{2
H\dot{\phi}^2}\, ,
\end{equation}
which for the function $f(R)$ and for the Hubble rate of Eqs.
(\ref{rsquaremodelfr}) and (\ref{hubblerate}), the second slow-roll
index is exactly equal to zero. Note that by using Eq.
(\ref{epsilon2general}), we obtain not a leading order result, but
an exact result for $\epsilon_2$. If we calculate $\epsilon_2$ by
using the solution $\phi (t)$ appearing in Eq. (\ref{phioft}), we
would obtain a leading order result, so it would be less accurate.
Let us proceed to the rest of the slow-roll indices appearing in
(\ref{slowrollgenerarlfrphi}), and we start off with $\epsilon_1$
 and $\epsilon_3$, which are found to be equal to,
\begin{equation}\label{epsilon1analytic}
\epsilon_1=\frac{H_i}{(H_0-H_i
t)^2},\,\,\,\epsilon_3=-\frac{H_i}{(H_0-H_i t)^2}\, ,
\end{equation}
and accordingly, the function $E$ is equal to $E=48 \alpha (H_0-H_i
t)^2$, so the slow-roll parameter $\epsilon_4$ is equal to
$\epsilon_3=\epsilon_4$. Since all the slow-roll parameters satisfy
the following condition $\epsilon_i\ll 1$, the spectral index of the
primordial curvature perturbations can be calculated by the
following formula,
\begin{equation}\label{ns1}
n_s\simeq 1-4\epsilon_1-2\epsilon_2+2\epsilon_3-2\epsilon_4\, ,
\end{equation}
so by substituting the resulting forms of the slow-roll indices, we
obtain,
\begin{equation}\label{nstimdependent}
n_s\simeq 1-\frac{4 H_i}{(H_0-H_i t)^2}\, ,
\end{equation}
which is a function of the cosmic time. Accordingly, the function
$Q_s$ appearing in Eq. (\ref{qsfunction}), is equal to,
\begin{equation}\label{qstimedependece}
Q_s=\frac{96 \alpha  H_i^2}{(H_0-H_i t)^2
\left(1-\frac{H_i}{(H_0-H_i t)^2}\right)^2}\, .
\end{equation}
The scalar-to-tensor ratio is equal to \cite{Nojiri:2017ncd},
\begin{equation}\label{scalartotensorratio}
r=\frac{8 Q_s}{F}\, ,
\end{equation}
so by substituting (\ref{qstimedependece}), we finally get,
\begin{equation}\label{scalartotensorratiotimedep}
r=\frac{32 H_i^2}{(H_0-H_i t)^4 \left(1-\frac{H_i}{(H_0-H_i
t)^2}\right)^2}\, .
\end{equation}
It is more convenient to express the cosmic time as a function of
the $e$-foldings number, and since we have the Hubble rate given,
this can easily be done. Indeed, from the Hubble rate
(\ref{hubblerate}) and also by the condition $\epsilon_1(t_f)\sim
1$, where $t_f$ is the time instance that inflation ends, we obtain
\begin{equation}\label{tffinalvalues}
t_f\simeq \frac{H_0}{H_i}\, .
\end{equation}
Also by using the definition of the $e$-foldings number,
\begin{equation}\label{efoldings}
N=\int_{t_k}^{t_f}H(t)\mathrm{d} t\, ,
\end{equation}
and by setting $t_k\ll t_f$, we obtain,
\begin{equation}\label{asxet1}
N\simeq \frac{H_0^2}{2H_i}\, ,
\end{equation}
By combining (\ref{tffinalvalues}) and (\ref{asxet1}) we get the
approximate relation between the $e$-foldings and the time instance
$t_f$, which is,
\begin{equation}\label{tfn}
t_f\simeq \frac{2N}{H_0}\, .
\end{equation}
Hence, by calculating the spectral index of primordial curvature
perturbations $n_s$ (\ref{nstimdependent}) and also the
scalar-to-tensor ratio (\ref{scalartotensorratiotimedep}), at the
time instance $t_f$ and also by taking the large-$N$ limit, we
obtain the observational indices,
\begin{equation}\label{nsandrfinalresults}
n_s\simeq 1-\frac{2}{N},\,\,\,r\simeq \frac{8}{N^2}\, .
\end{equation}
As it can be seen in Eq. (\ref{nsandrfinalresults}), the spectral
index is identical to the $R^2$ inflation case, however the
scalar-to-tensor ratio is not the same, since in the $R^2$ case it
is $r\sim \frac{12}{N^2}$.
\begin{figure}[h]
\centering
\includegraphics[width=18pc]{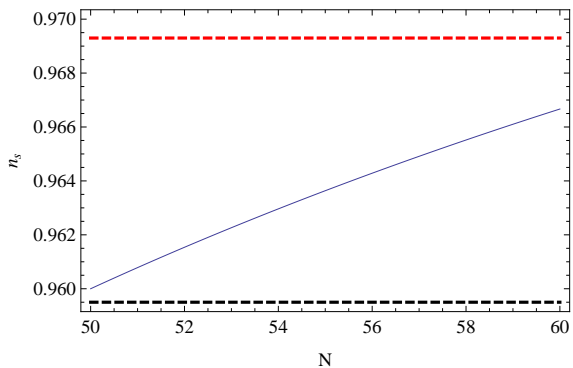}
\caption{The spectral index $n_s=1-\frac{2}{N}$ as a function of the
$e$-foldings number $N$, for $N=[50,60]$. The red dashed line
corresponds to the upper bound specified by the 2015-Planck data for
the spectral index $n_s=0.9693$, while the black dashed line is the
lower bound of the 2015-Planck data $n_s=0.9595$.}\label{plot1}
\end{figure}
Hence, in the case at hand, the scalar-to-tensor ratio is smaller in
magnitude. A striking feature is that the resulting observational
indices are $\alpha$-independent, however this is possibly due to
the model we used. In order to have a quantitative idea of the
predictability of the model, let us give some characteristic values,
for example if we choose $N=60$, the spectral index becomes
$n_s\simeq 0.966 $, and by direct comparison with the Planck data
\cite{Ade:2015lrj},
\begin{equation}
\label{planckdata} n_s=0.9644\pm 0.0049\, , \quad r<0.10\, ,
\end{equation}
this means that it is within the allowed range of values for $n_s$,
which are $n_s=[0.9595,0.9693]$. Also the scalar-to-tensor ratio for
$N=60$ is $r=0.00222$, while the predicted scalar-to-tensor ratio
for the $R^2$ model is $r\simeq 0.0033333$, which is larger. In Fig.
\ref{plot1} we plot the behavior of the spectral index as a function
of $N$ and in Fig. \ref{plot2}, we plot the scalar-to-tensor ratio
for the modified $R^2$ model with scalar field (red curve) and for
the $R^2$ model (blue curve). As it can be seen, in both cases, the
values of the observational indices are well within the
observational constraints of the Planck collaboration
\cite{Ade:2015lrj} but also within the allowed values of the
BICEP2/Keck-Array data \cite{Array:2015xqh}, which constrain the
scalar-to-tensor ratio as follows,
\begin{equation}
\label{scalartotensorbicep2} r<0.07\, ,
\end{equation}
Hence, as we demonstrated in this section, it is possible to obtain
a viable slow-roll quasi-de Sitter evolution from an $f(R)$ gravity
which is not viable in vacuum, however in the presence of a scalar
field with appropriately chosen potential, this theory becomes
viable. In principle, other choice of $f(R)$ gravity may be chosen,
however if the $f(R)$ gravity has a complicated functional form,
analyticity is difficult to achieve and consequential approximations
might be needed.

\section{Conclusions}

In this paper we studied the simplest form of an $f(R,\phi)$
gravity, and specifically an $f(R,\phi)$ gravity in the presence of
a canonical scalar field. We focused on the inflationary era and we
assumed that the $f(R)$ gravity has a specific form, leaving the
scalar field theory free. Then, we fixed the Hubble rate of the
cosmological evolution and we investigated which scalar potential
can reproduce such a cosmological evolution in combination with the
$f(R)$ gravity. To this end, we assumed for both the $f(R)$ and the
scalar theory, the slow-roll conditions apply, that is
$\ddot{\phi}\ll H\dot{\phi}$, for the scalar field, and also
$\ddot{H}\ll H\dot{H}$ and $\dot{H}\ll H^2$ for the $f(R)$ frame.
Our calculations showed that the $R^2$ inflation in the presence of
a scalar field leads to inconsistencies. So we introduced a variant
form of the $R^2$ inflation, namely $f(R)=R-\alpha R^2$, and we
found the potential of the scalar field that in conjunction with the
modified $R^2$ gravity, realizes the quasi-de Sitter inflationary
evolution. In addition, the value of the parameter $\alpha$ was
determined by using the slow-roll condition on the scalar field. For
the resulting theory we calculated in detail the spectral index of
primordial curvature perturbations and the scalar-to-tensor ration
at leading order in the $e$-foldings number $N$, and as we
demonstrated, compatibility with the Planck data but also with the
BICEP2/Keck-Array data was achieved. Remarkably, the resulting
theory is independent of the parameter $\alpha$, at least at leading
order in $N$.

An interesting issue which we did not address, but would be
certainly interesting, is the study of the reheating era. This would
be particularly interesting, since the presence of the scalar field
can assist the reheating era, so this may directly affect the
dynamics of the reheating era, and in effect may directly modify the
reheating temperature. We hope to address this issue in a future
work.

In principle, other choices for the Hubble rate and the $f(R)$
gravity can be made, in order to obtain a viable inflationary
cosmology. In this paper we made the simplest choices in order to
obtain analytic results, at least in most of the cases.

\section*{Acknowledgments}

Financial support by the Research
Committee of the Technological Education Institute of Central Macedonia, Serres,
under grant SAT/ME/170118-14/14, is gratefully acknowledged.

\end{document}